\documentclass[referee]{cjaa}
\include{graphicx}

\begin{document}

\title{Re-Identification of the `Enigmatic' X-ray Source 1RXS~J114003.0+124112}

\author{Jianghua Wu
    \inst{1}\mailto{}
  \and Tigran Movsessian
    \inst{2}
  \and Yang Chen
    \inst{3}
  \and Xiangtao He
    \inst{3}
  \and Xu Zhou
    \inst{1}
  \and Jun Ma
    \inst{1}
}

\offprints{Jianghua Wu}

\institute{National Astronomical Observatories, Chinese Academy of Sciences,
           Beijing 100012, China
          \email{jhwu@bao.ac.cn}
    \and
           Byurakan Astrophysical Observatory, Aragatsotn prov. 378433, Armenia
    \and
           Department of Astronomy, Beijing Normal University, Beijing 100875,
           China
}

\date{Received; Accepted}

\abstract{The {\sl ROSAT} X-ray source 1RXS~J114003.0+124112 was identified as
a starburst galaxy at redshift 0.177 by He et al. (\cite{he01}). Meanwhile,
the authors also noted that the source is almost two orders of magnitude
brighter in X-ray than the X-ray-brightest starburst galaxy and it seems to
be in a merging system, making this source an enigmatic system for further
observations. This paper reports the re-identification of
1RXS~J114003.0+124112 with the observations on the 2.6 m telescope at
Byurakan Astrophysical Observatory, Armenia and with the SDSS data. The
results indicate that the starburst activity is associated with the brighter
object in the system, while the fainter object is a typical Seyfert 1 galaxy
at redshift 0.282. Therefore, the two objects are not in a merging
system, and the Seyfert 1 galaxy naturally accounts for the high X-ray flux.
Three more objects reside in the vicinity, but they are all too faint to be
responsible for the high X-ray flux.
}

\authorrunning{Wu et al.}
\titlerunning{Re-Identification of 1RXS~J114003.0+124112}

\maketitle

\section{Introduction}
In the first paper of the Multi-Wavelength Quasar Survey (MWQS, He et al.
\cite{he01}), the {\sl ROSAT} source 1RXS~J114003.0+124112 was identified as
a starburst galaxy at redshift 0.177 based on its optical spectrum. At the
mean time, the authors have also noticed its peculiarities as a starburst
galaxy: Its X-ray luminosity is almost two orders of magnitude higher than
those of the X-ray-brightest starbursts. When assuming a power-law with a
photo index of $\Gamma=2.2$ and adopting a Galactic neutral hydrogen column
density of $3.15\times10^{20}\,\rm{cm^{-2}}$ (computed by using Colden: the
Galactic Neutral Hydrogen Density Calculator,
http://cxc.harvard.edu/toolkit/colden.jsp), its X-ray flux is $0.67\times10
^{-12}\,\rm{erg\,s^{-1}\,cm^{-2}}$, and the luminosity is $9.78\times
10^{43}\,\rm{erg\,s^{-1}}$ when adopting $H_0=50\,\rm{km\,s^{-1}\,Mpc^{-1}}$
and $q_0=0.5$. The X-ray luminosity is typical of Seyfert 1 galaxies. In
addition, the source shows on the
Digitized Sky Survey (DSS) images two closely separated nuclei, one being
brighter and bluer, and the other being fainter and redder. They are
surrounded and connected by some fuzzy structures, implying a merging system.
From these results three questions emanate immediately: (1) On which nucleus
has the spectrum been taken? It is likely on the brighter one, but it needs
to be confirmed. (2) How can we understand the high X-ray luminosity from
this system? A starburst galaxy is unlikely able to emit such a high X-ray
luminosity. (3) What's the role of the merging in this system? Is it related
to the high X-ray luminosity?


In order to answer these questions, we re-observed this source with the 2.6
m telescope at Byurakan Astrophysical Observatory, Byurakan, Armenia. Soon
afterwards, we found that the spectrum of the fainter object has also been
obtained by the Sloan Digital Sky Survey (SDSS) and released on the website.
Here we present the results of these new observations on this enigmatic X-ray
source.

\begin{figure}
\includegraphics[width=6.6cm]{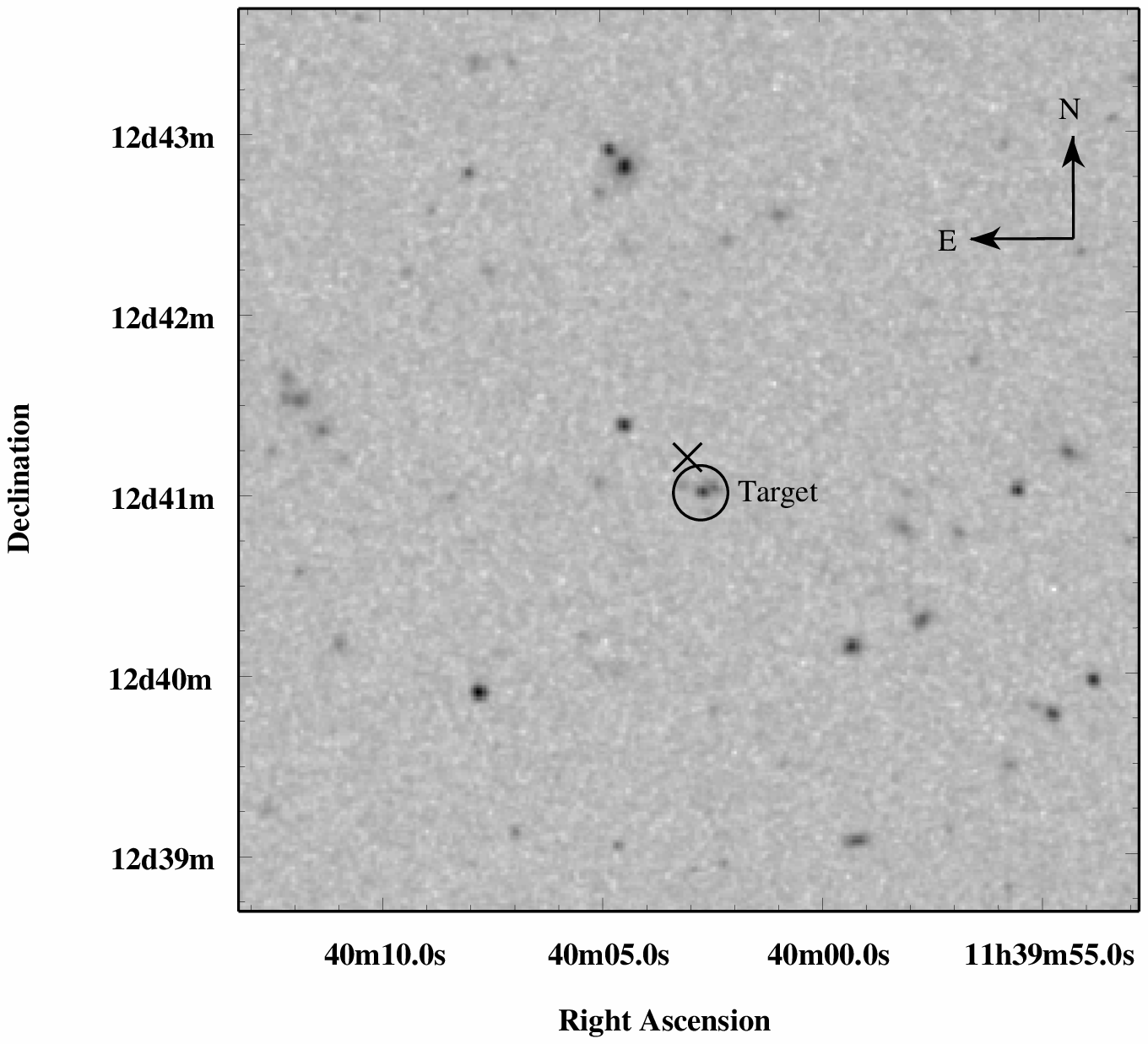}
\includegraphics[width=6.6cm]{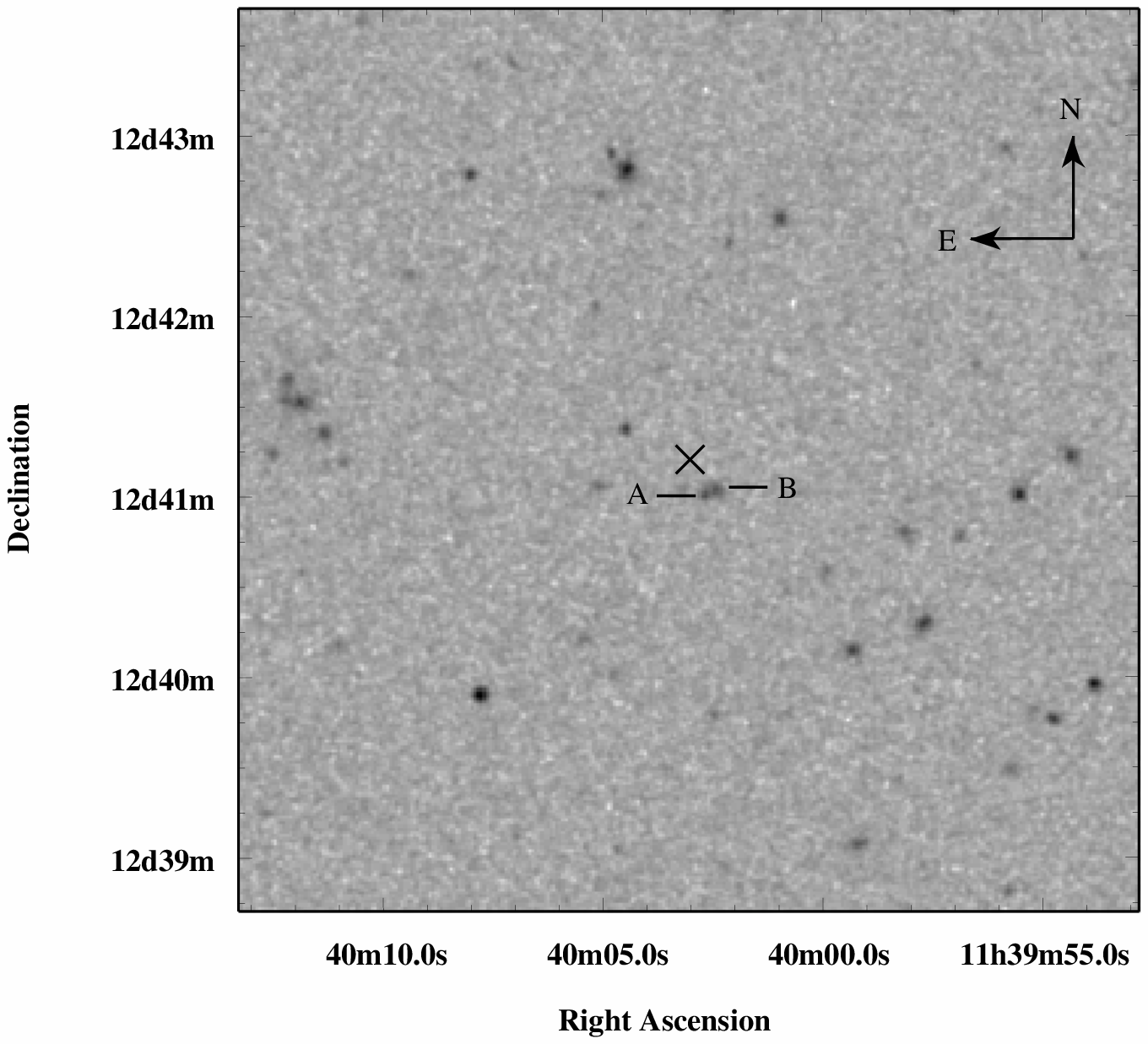}
\caption{The DSS blue (left) and red (right) images centered on the X-ray
position. The target is labeled by a circle for spectroscopic observations,
while the X-ray position is labeled by a cross. Two nuclei are connected and
surrounded by some fuzzy structures. At the east and south of them, there
are two more and fainter objects. The sizes are both $5'\times5'$.}
\label{F1}
\end{figure}

\section{The Images}

\subsection{The DSS Images}

Fig.~\ref{F1} presents the DSS blue (left) and red (right) images of the
system centered on the {\sl ROSAT} X-ray position which is marked by a cross
 The system is labeled by a circle on the blue image. There is a bright
nucleus (Object A) near the center of the circle. A much fainter nucleus
(Object B) resides to its close northwest. On the red image, however, they
have comparative brightness, implying that Object B is relatively redder than
Object A. The separation between the two nuclei is very small, and they are
apparently surrounded and connected by some fuzzy structures, implying a
merging system. The starburst spectrum in He et al. (\cite{he01}) should
have focused on Object A, and may have included some light from Object B as
well. There are two more even fainter objects that are located to the east
and south of Objects A and B on the blue image, but they can hardly observed
on the red image. They are too faint to be responsible for the high X-ray
luminosity.


\begin{figure}
\includegraphics[width=6.6cm]{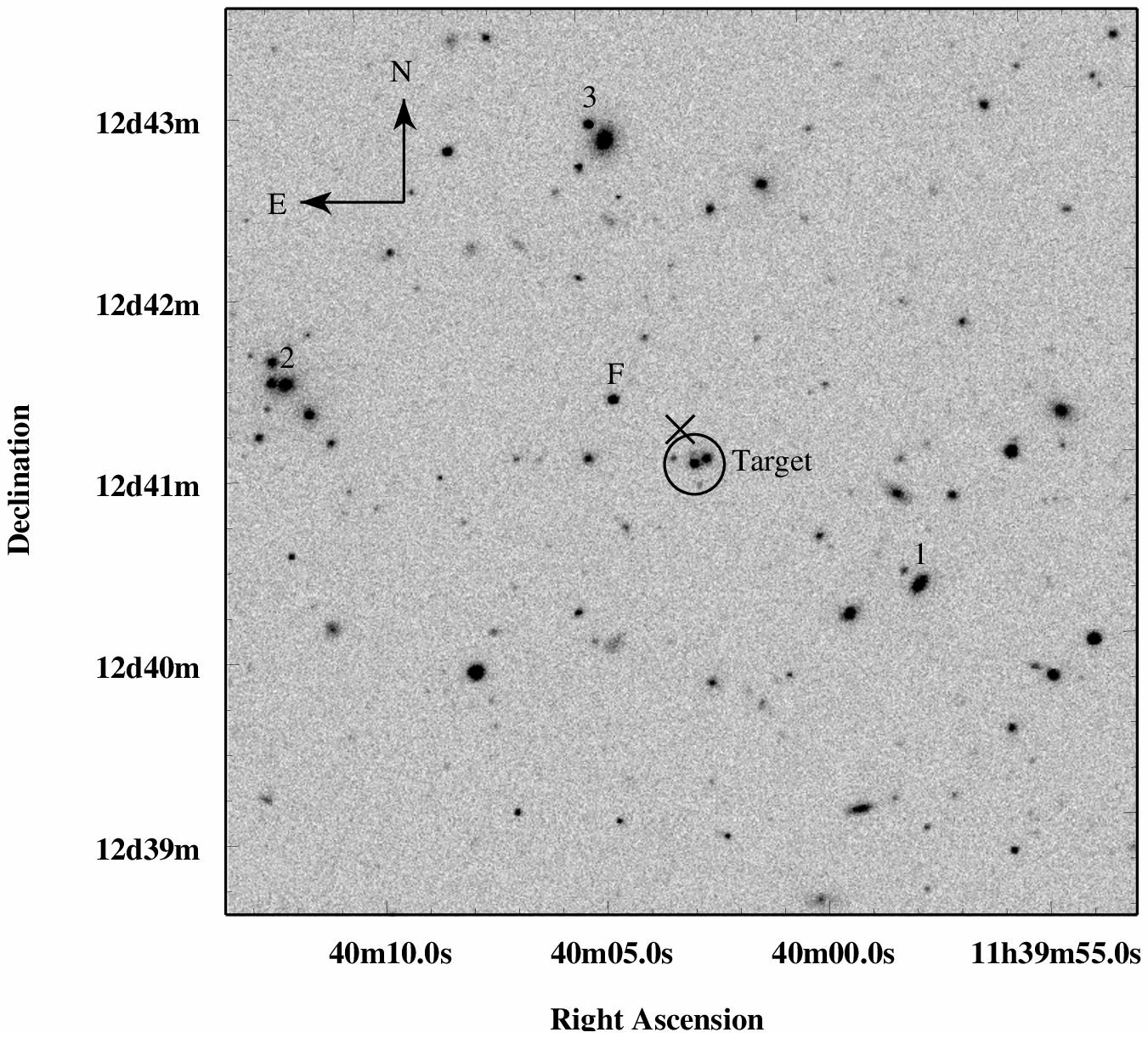}
\includegraphics[width=6.6cm]{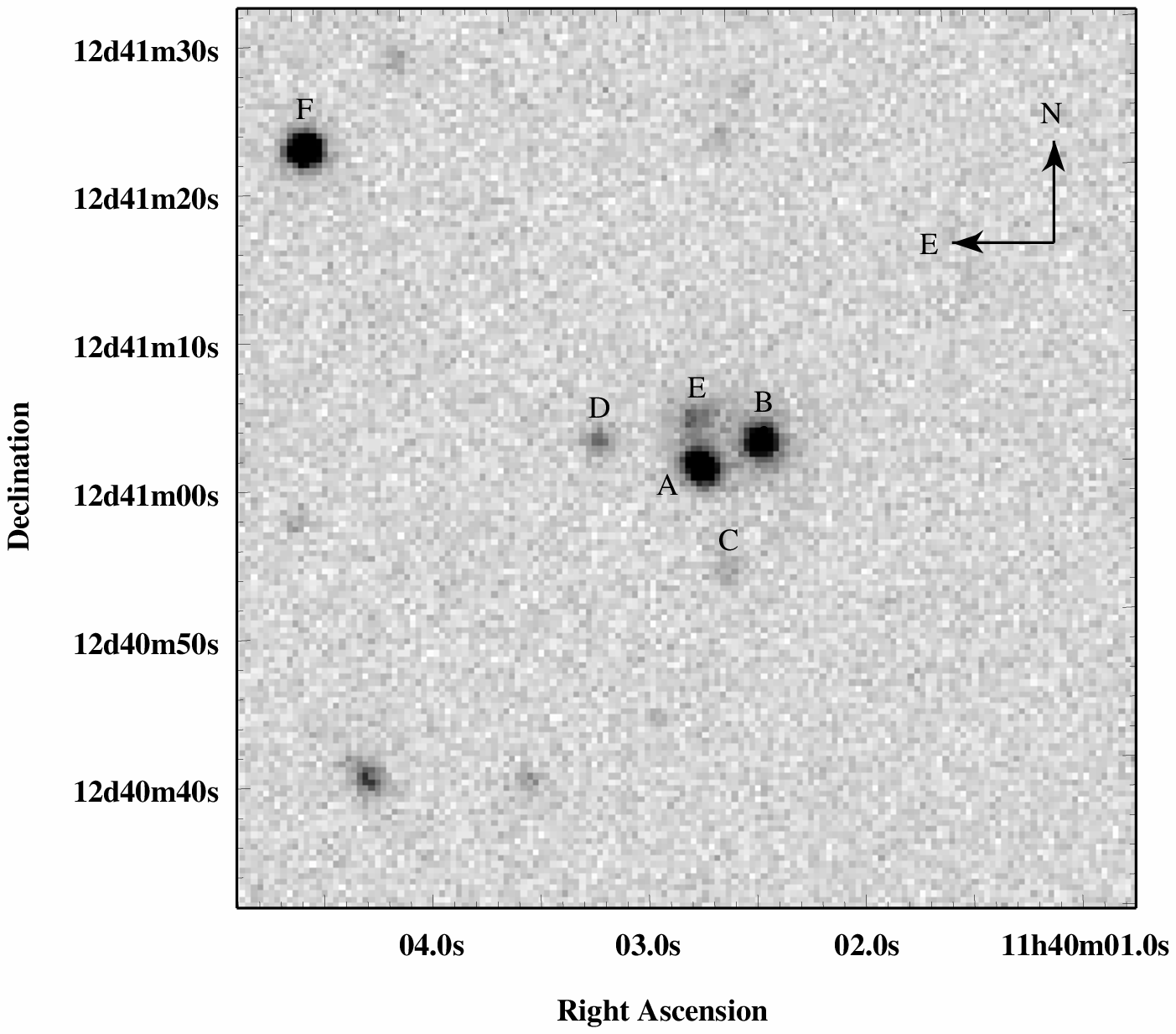}
\caption{The SDSS images in the $r'$ (left) and $i'$ (right) bands centered
on Object A. The sizes are $5'\times5'$ and $1'\times1'$, respectively. The
circle labels the target for spectroscopic observations, while the cross
marks the X-ray position. Objects 1 and 2 are two galaxies both at redshift
0.274, and Object 3 is a quasar at $z=2.482$, according to their SDSS spectra.}
\label{F2}
\end{figure}

\subsection{The SDSS Images}

The system is well resolved on the SDSS images. Fig.~\ref{F2} illustrates
the SDSS $r'$ (left) and $i'$ (right) band images in $5'\times5'$ and
$1'\times1'$, respectively. They are both centered on Object A, and the cross
marks the X-ray position. No clear physical connections can be seen between
Objects A and B. The projected separation between them is $\sim4\farcs4$,
as measured from the SDSS images. Three more and much fainter objects, C,
D, and E, are also observable, among which Objects C and D show up on the
DSS blue image, too (see Fig.~\ref{F1},{\sl left}). Object E is not
observable on the DSS images. It is much brighter in the $i'$ and $z'$ bands
than in the other three SDSS bands (see Table~\ref{T1}). As Objects C and D,
it is also too faint to be responsible for the high X-ray luminosity.

To the northeast of the X-ray position there is a bright object, Object F.
Although it is bright and shows a blue color on the SDSS images, it is not
isolated as an AGN candidate by the target selection algorithm of the SDSS
program (Richards et al. \cite{rich02}). So it is not likely to be an AGN
and will not be responsible for the high X-ray luminosity.


\section{The Spectroscopic Observations and Data Reduction}

Observations were carried out on 2005 June 12 with the 2.6 m telescope at
Byurakan Astrophysical Observatory, using SCORPIO (Afanasiev \& Moiseev
\cite{afana05}) spectral camera attached at the prime focus of the telescope.
The multi-mode camera was used in long-slit mode, and during our set the
slit length was $7\arcmin$ and width was $2\arcsec$. The slit is fixed
in E-W direction and passed two objects simultaneously. The instrument was
equipped with a $600\,\rm{g\,mm^{-1}}$ grism, providing spectral range
$3900-7200\,\rm{\AA}$ with a $6\,\rm{\AA}$ resolution. A detector Loral
$2063\times2058$ pixel CCD matrix was used, which provides $14'\times14'$
field in imagery mode and $3300\,\rm{\AA}$ spectral range in long slit mode.

\begin{figure}
\hspace{0.7cm}
\includegraphics[width=6.15cm]{fig3.a.ps}
\hspace{0.8cm}
\includegraphics[width=6.15cm]{fig3.b.ps}
\includegraphics[width=7cm]{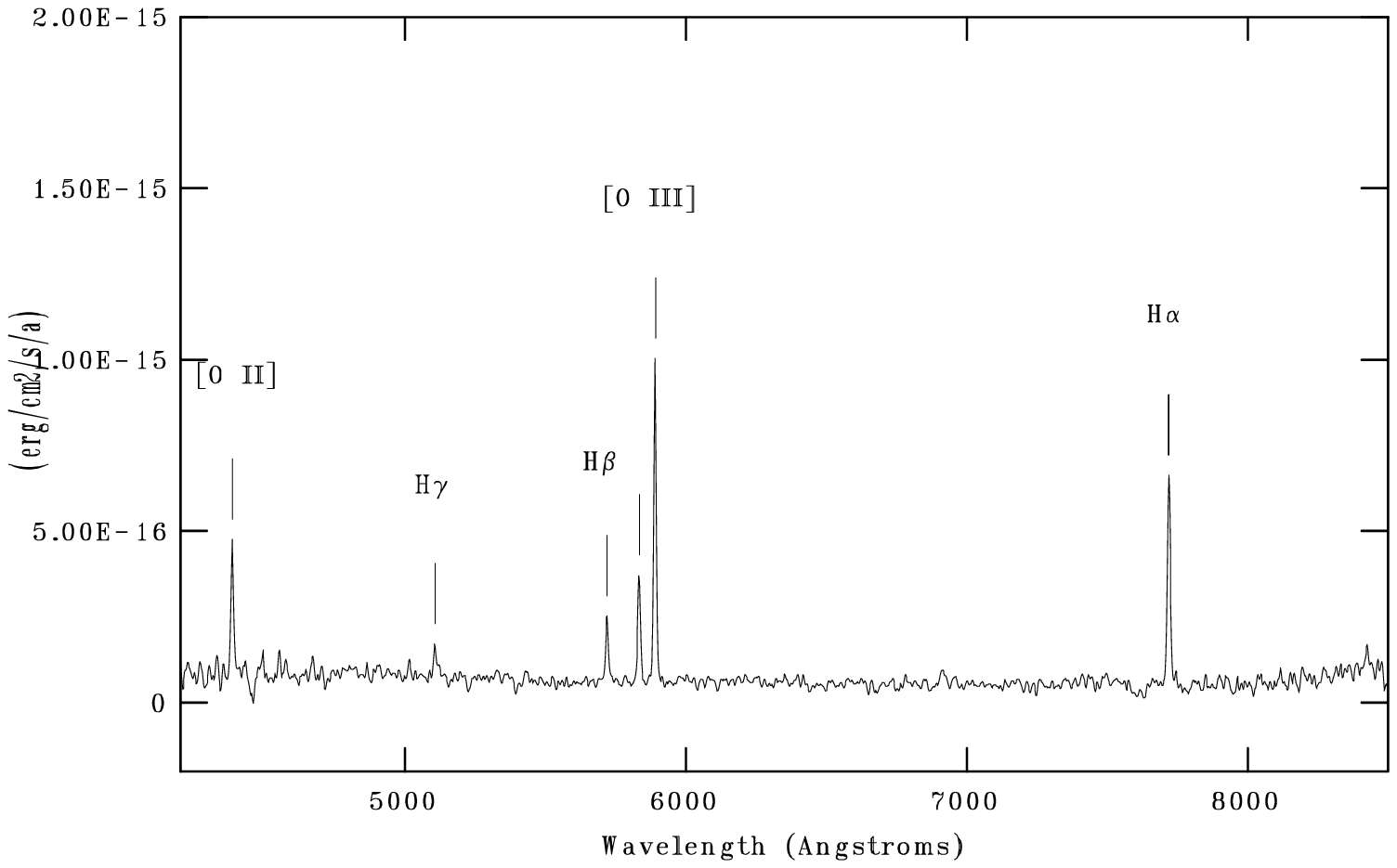}
\includegraphics[width=7cm]{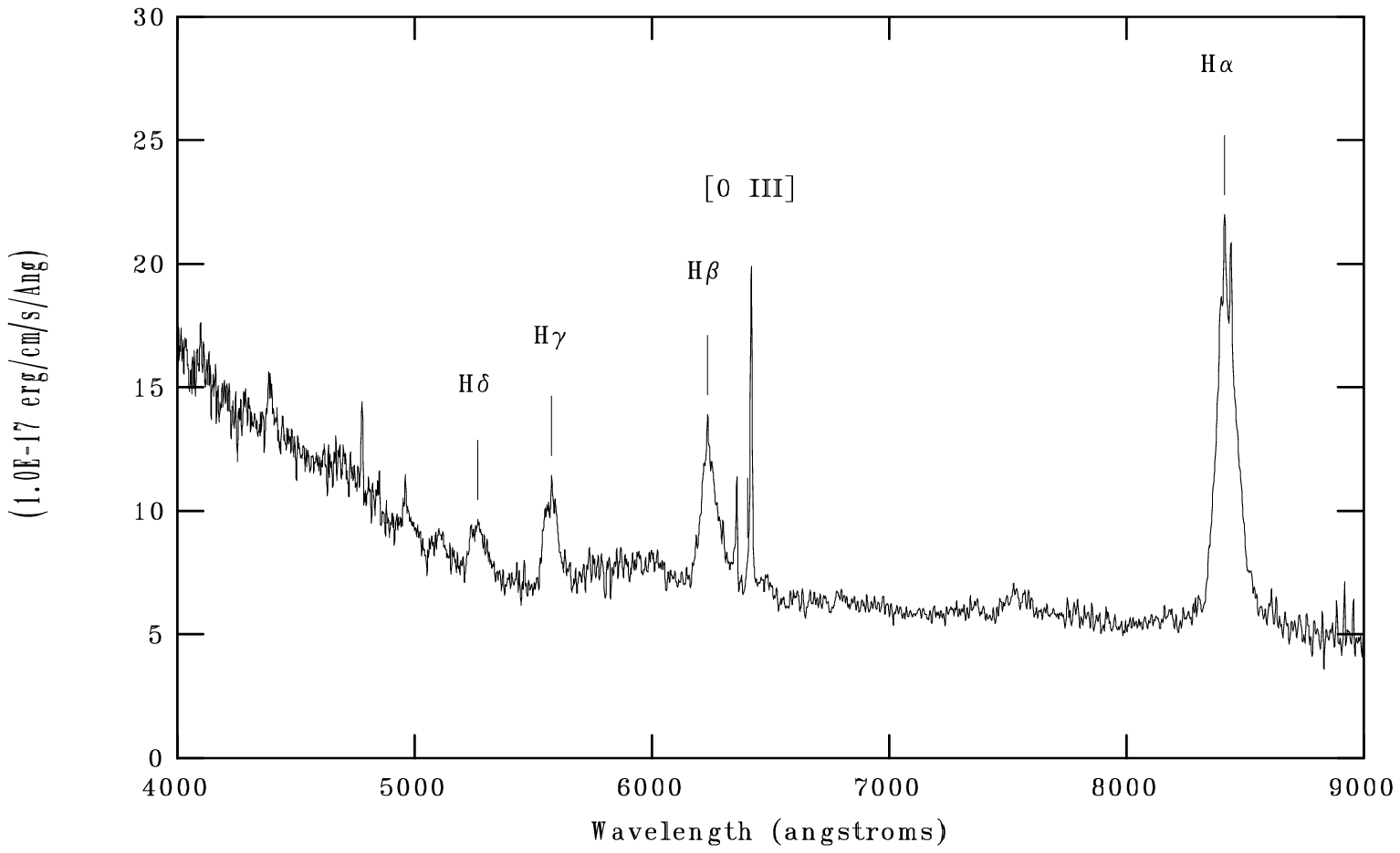}
\caption{Spectra of Objects A and B. The upper two spectra are taken on the
2.6 m telescope at Byurakan Astrophysical Observatory (flux is in arbitrary
scale). The broad trough at 7600 {\AA} is night sky absorption in Object B.
The spectrum of
Object A taken on the 2.1 m telescope at KPNO and the SDSS spectrum of Object
B are also shown on the lower-left and lower-right respectively for a
comparison. The common emission lines are labeled. Object A has a typical
spectrum of starburst, while Object B has a typical spectrum of Type 1 AGNs.
The redshifts are 0.177 and 0.282, respectively.}
\label{F3}
\end{figure}

Data reduction was done by using IRAF with the standard procedures, including
bias and flat field corrections, 1-D spectra extraction, and wavelength
calibration. The spectra of two objects are displayed in Fig.~\ref{F3}.

\section{The Spectra}

The spectra taken on the 2.6 m telescope at Byurakan Astrophysical Observatory
are shown in Fig.~\ref{F3} (upper). The results confirm that Object A is a
starburst galaxy at redshift $0.177\pm0.006$. Also presented is the spectrum
taken on the 2.1 m telescope at the Kitt Peak National Observatory (KPNO),
USA on 1999 September 25. It shows a perfect power-law continuum and some
quite narrow emission lines. Moreover, the spectrum displays some extreme
properties for a starburst galaxy: (1) It shows a nearly unreddened Balmer
decrement; (2) It has currently undetected [N~II] emission, placing it at an
extreme position for starburst or H~II galaxies in the diagnostic diagrams
of Veilleux \& Osterbrock (\cite{veilleux87}). So it might be interesting to
make further observations on this starburst galaxy.

Object B is a Seyfert 1 galaxy at redshift 0.282 based on the spectroscopic
observation on the 2.6 m telescope (see upper-right of Fig.~\ref{F3}). At the
same time, it should have been isolated as an AGN candidate for spectroscopic
observation by the target selection algorithm of the SDSS (Richards et al.
\cite{rich02}), so its spectrum was also obtained by the SDSS program on 2004
March 19. The lower-right of Fig.~\ref{F3} gives its SDSS spectrum. The two
spectra of Object B have some differences in continuum, i.e., slightly
different slopes and the night sky absorption at 7600 {\AA} is not subtracted
satisfactorily in the upper spectrum. There should be resulted from the
different observing facilities and different data reduction methods. But the
emissions in both spectra are exactly at the same wavelengths, and both spectra
indicate that Object B is typical for a Type 1 AGN, with blue color, broad
and strong permitted emission lines and Fe~II lines, and narrow forbidden
lines, which give a redshift of $0.282\pm0.001$. Therefore, Object B
cannot form a merging system
with Object A, which has a redshift of 0.177. Moreover, the `enigmatic' high
X-ray luminosity is naturally explained, it should be dominated by the emission
from Object B, the Seyfert 1 galaxy, rather than from Object A, the starburst
galaxy.

At the new redshift, the X-ray luminosity should be re-calculated as
$2.59\times10^{44}\,\rm{erg\,s^{-1}}$.

The properties of all five objects are summarised in Table~\ref{T1}. The
first column is the Object ID. The second and the third columns are the
positions of the objects. The following five columns are the magnitudes in
five SDSS bands. The 9th and 10th columns are redshift and classification
of the objects (SB signifies starburst, and S1, Seyfert 1 galaxy).

\begin{table}
\caption{Properties of All Five Objects}
\begin{tabular}{cccccccccc}
\hline\hline
Obj &  R.A. & DEC & $u'$ & $g'$ & $r'$ & $i'$ & $z'$ & $z$ & Type \\
    & (J2000.0) & (J2000.0) & (mag) & (mag) & (mag) & (mag) & (mag) & & \\
\hline
A & 11:40:02.69 & 12:41:00.67 & 19.40 & 19.04 & 18.66 & 18.69 & 18.83 & $0.177\pm0.006$ & SB \\
B & 11:40:02.41 & 12:41:02.10 & 19.20 & 19.27 & 18.71 & 18.48 & 17.85 & $0.282\pm0.001$ & S1 \\
C & 11:40:02.58 & 12:40:53.28 & 21.82 & 21.55 & 21.18 & 21.12 & 21.38 &   &  \\
D & 11:40:03.15 & 12:41:02.51 & 21.52 & 22.64 & 21.04 & 20.63 & 20.47 &   &  \\
E & 11:40:02.70 & 12:41:03.84 & 24.89 & 24.34 & 24.78 & 20.45 & 19.51 &   &  \\
\hline
\end{tabular}
\label{T1}
\end{table}

\section{Conclusions and Discussions}

The `enigmatic' X-ray source, 1RXS~J114003.0+124112, is re-identified with
the observations on the 2.6 m telescope at the Byurakan Astrophysical
Observatory, Armenia and with the SDSS data. The starburst activity is
proved to be associated with the brighter object in the system, while the
fainter object is identified as a typical Seyfert 1 galaxy at redshift
0.282. Therefore, the two objects do not form a merging system, and the
Seyfert 1 galaxy naturally accounts for the high X-ray flux. Three more
objects reside in the vicinity, but they are too faint to be responsible
for the high X-ray flux.

We once suspected that this source was a so-called starburst/Seyfert composite
galaxy, a peculiar sub-class of AGNs that were discovered by cross-correlation
of the {\sl ROSAT} All-Sky Survey with the {\sl IRAS} Point Source Catalog
(Moran et al. \cite{moran96}). This class of objects show optical spectra
characteristic of starbursts based on the quite narrow emission lines and the
diagnostic diagrams (Veilleux \& Osterbrock \cite{veilleux87}), while their
X-ray luminosities are around $10^{42-43}\,{\rm erg\,s^{-1}}$, which is
typical of Seyfert galaxies, and which is one or two orders of magnitude
higher than those of the brightest starburst galaxies (Georgantopoulos
\cite{georg00}; Pappa et al. \cite{pappa02}; Georgantopoulos, Zezas,
\& Ward \cite{georg03}; Georgantopoulos et al. \cite{georg04}). The new
spectroscopic observations ruled out this suspicion.

There may be a cluster of galaxies towards the direction of this system.
On the SDSS $r'$ band image (Fig.~\ref{F2}, {\sl left}), Objects 1 and 2
are two galaxies both at redshift 0.274, very close to the redshift of
Object B. Object 2 itself may belong to a subgroup of galaxies. Moreover,
there is a clear overdensity of extended objects scattered in and slight
outside of this image. Most of them have similar colors, implying similar
distances. A photometric redshift estimates of these objects may help to
confirm the existence of the cluster.

\begin{acknowledgements}
The authors thank the anonymous referee for insightful comments.
The Digitized Sky Surveys were produced at the Space Telescope Science
Institute under U.S. Government grant NAG W-2166. The images of these
surveys are based on photographic data obtained using the Oschin Schmidt
Telescope on Palomar Mountain and the UK Schmidt Telescope. This work has
made use of the images and spectrum from the Sloan Digital Sky Survey
(SDSS, http://www.sdss.org). This work has been supported by the Chinese
National Natural Science Foundation, No. 10473012, 10573020, and 10303003.
\end{acknowledgements}

\end{document}